# Ultrahigh ion diffusion in oxide crystal by engineering the interfacial transporter channels

Liang Li[1+], Min Hu[2+], Changlong Hu[1], Bowen Li[1], Shanguang Zhao[1], Guobin Zhang[1], Liangbin Li[1], Jun Jiang[2], Chongwen Zou[1*]

[1] *National Synchrotron Radiation Laboratory, School of Nuclear Science and Technology, University of Science and Technology of China, Hefei, Anhui 230029, P. R. China*

[2] *Hefei National Laboratory for Physical Sciences at the Microscale, Collaborative Innovation Center of Chemistry for Energy Materials, CAS Center for Excellence in Nanoscience, School of Chemistry and Materials Science, University of Science and Technology of China, Hefei, Anhui 230026, P. R. China*

+These two authors contributed equally to this paper.

*Corresponding Author: czou@ustc.edu.cn

# Abstract

The mass storage and removal in solid conductors always played vital role on the technological applications such as modern batteries, permeation membranes and neuronal computations, which were seriously lying on the ion diffusion and kinetics in bulk lattice. However, the ions transport was kinetically limited by the low diffusional process, which made it a challenge to fabricate applicable conductors with high electronic and ionic conductivities at room temperature. It was known that at essentially all interfaces, the existed space charge layers could modify the charge transport, storage and transfer properties. Thus, in the current study, we proposed an acid solution/$WO_3$/ITO structure and achieved an ultrafast hydrogen transport in $WO_3$ layer by interfacial job-sharing diffusion. In this sandwich structure, the transport pathways of the protons and electrons were spatially separated in acid solution and ITO layer respectively, resulting the pronounced increasing of effective hydrogen diffusion coefficient ($D_{eff}$) up to $10^6$ times. The experiment and theory simulations also revealed that this accelerated hydrogen transport based on the interfacial job-sharing diffusion was universal and could be extended to other ions and oxide materials as well, which would potentially stimulate systematic studies on ultrafast mixed conductors or faster solid-state electrochemical switching devices in the future.

## Introduction

The element doping and ion migration in solid oxide compounds could regulate the properties of functional materials(*1-3*) ,which always played vital role on various applications, such as electrochromic windows(*4-7*), electrical tri-state phase transition(*8, 9*), selective electrocatalysis(*10*) , surface self-assembling(*11*), ion storage(*12*) and synaptic transistors(*13, 14*). Naturally, as a key process in material and physical science(*15-17*), the migration or diffusion of atoms in solid, was also significant to be investigated for more tunable and high-performance devices(*18-23*). Quick transport of atoms doping in films could shorten the dynamic process to improve the performance of devices effectively(*22, 24*). This kinetic process was characterized by a key parameter, chemical diffusion coefficient ($D_{eff}$), which was very important in the field of diffusion in solids(*25, 26*).

Normally the ions transport was kinetically limited by the extremely low diffusional process driven by the concentration gradient or even by external electric field, which made it a challenge to fabricate applicable conductors with high electronic and ionic conductivities at room temperature. It had been suggested that at essentially all interfaces, the existed space charge layers could modify the charge transport, storage and transfer properties (*25, 26*). The recent reports showed that quick diffusion of lithium(*27*) and other alkali elements(*16*) had been realized in multilayer graphene, owing to its fast-track between layers and the collective effects(*16*). Another inspiring progress was the realization of ultrafast Ag storage and removal in the interface of $RbAg_4I_5$-graphite due to the chemical diffusion along the interface(*26*), which resulted the diffusion coefficient of Ag atoms along this interface up to $D_{eff} \sim 10^{-4}$ cm$^2$/s.

However, it was still a challenge to realize an ultrafast atomic migration in bulk solid, such as oxide solid materials. Though it had been reported that in oxides films, external energy input, such as applying electrostatic(*21*) or static magnetic(*23*) field, was a useful way to realize tunable mass transport. This route could change the initial and final energy state through inducing external field, as shown in Supplementary Fig. S1A. To accelerate the atomic diffusion process without external energy input, reducing the diffusion activation energy ($E_a$) was the possible way according to the Fick's laws of diffusion and the Arrhenius equation of diffusion coefficient:

$$D \propto exp(-E_a/k_BT),$$

where $k_B$ was the Boltzmann constant and $T$ was the surrounding temperature(*16, 28*). The schematic evolution of energy barrier was show in Supplementary Fig. S1B. According to this strategy, many attempts had been made in oxide material to improve the $D_{eff}$ value with low $E_a$ by lattice defect engineering(*18-20, 22, 24, 29*), while the final increase was not such pronounced. For example, by increasing the density of domain boundaries in $VO_2$ crystal as the "highway", the diffusion coefficient of H atoms only improved by about twenty times(*22*).

It was known that the atomic diffusion consisted of cationic and electronic migration, which could be described by the concept of ambipolar diffusion(*30*) and was successfully applied in plasma physics and astrophysics(*31*). From this viewpoint, it was possible to achieve ultrafast diffusion of doped atoms along an artificial hetero-junctions or interfaces in multi-phase systems. This fabricated interface composited of ion and electron conducting phases, which made the transport pathways of ions and electrons spatially separated and driven the transport of a neutral component via space charge effect.

Based on the above strategy, in the current study, we proposed an acid solution/$WO_3$/ITO structure and achieved an ultrafast hydrogen transport in $WO_3$ layer with the $D_{eff}$ value of about $10^{-1}$ $cm^2/s$, which was almost ~$10^6$ times higher than that in $WO_3$ bulk lattice. Interestingly, due to the distinct electrochromic property of $WO_3$, this ultrafast hydrogen transport could be directly observed by eyesight. The experiment and theory simulations also revealed that the accelerated hydrogen transport was mainly lying on the interfacial job-sharing diffusion of the protons and electrons in acid solution and ITO layer respectively. Furthermore, this proposed proton-electron synergistic diffusion mechanism was universal and could be extended to other ions and oxide materials as well, which would potentially stimulate systematic studies on ultrafast mixed conductors or other emerging ionic devices in the future.

## Accelerated diffusion of H atoms with proton and electron bridges

As a member of transition metal oxides, $WO_3$ was an ideal prototype to investigate the diffusion of doping atom in oxide films, owing to its electrochromic property by M (M = H, Li, Al etc.) intercalating. For example, H atoms could be easily driven into $WO_3$ film by external voltage gating, which was applicable for $WO_3$ based smart window device (Supplementary Fig. S2A). It was suggested that the reduction of $W^{6+}$ was responsible to H (or other M atoms) doping induced color change of $WO_3$, which was originated from the optical absorptions of small-polaron (*32, 33*). The detailed XRD and SEM/TEM investigations for the ~580nm amorphous $WO_3$ film before and after the H atoms insertion showed no obvious thickness and crystal structure changes (Supplementary Fig. S3). While due to color contrast between the initial transparent state and the H-doped blue state(*7*), the transport of H atom in $WO_3$ films could be directly visualized even by eyesight if the atomic diffusion was fast enough to be traced by the film color change. However, the concentration gradient induced dopants diffusion behavior in solid $WO_3$ lattice was always negligible, since the effective diffusion coefficients were extremely slow. For example, according to the references (*13-17*), the effective diffusion coefficient of H atoms in $WO_3$ bulk was only ~ $10^{-10}$ $cm^2/s$ in crystal and ~ $10^{-8}$ $cm^2/s$ in amorphous at room temperature (RT).

This extremely slow diffusion of H atoms in $WO_3$/sapphire film was directly reflected in the optical image of Fig. 1A. The blue region of the film was related to the heavily H doped $WO_3$ film ($H_xWO_3$), while the transparent region was the pure $WO_3$ film. It was observed that there existed a clear borderline between the deep-blue $H_xWO_3$ and transparent $WO_3$ area. Though a large H concentration gradient existed at the interface, almost no migration of the borderline was observed from the optical image after one day or even one week at ambient condition. Since the H atoms diffusion was directly associated with the color change of $WO_3$ film, the detailed transmittance-position curve plotted in Supplementary Fig. S4A showed that the borderline was almost kept at the same position even after one week, confirming this ignored H atoms diffusion in $WO_3$ film (*34*). It was reasonable if considering this ultra-low H atoms diffusion was just driven by the H concentration gradient at the $H_xWO_3$ -$WO_3$ interface lattice as shown Fig.1D. In fact, previous studies also demonstrated that the H atoms doping induced micro/nano patterns in $WO_3$ film showed no obvious diffusions at ambient condition, which was suitable for functional devices applications (*35*).

It was known that the neutral H atom diffusion could be separated by the proton and electron migrations, thus based on the $H_xWO_3/WO_3$ interface, we established a protonic bridge by covering a sulfuric acid layer (Fig. 1E), which showed a high proton conductivity. From the optical image in Fig. 1B, it was observed the considerable borderline migration within several minutes, which indicated that this protonic bridge could greatly accelerate the in-plane transport process of H atoms. The ultrafast immigration of H atoms in $WO_3$ layer was also examined by XPS in Supplementary Fig. S2C~S2F, confirmed the H atoms insertion in $WO_3$ layer and resulted the blue color change.

In addition, if we deposited similar $WO_3$ film on a conductive ITO-glass substrate, an electronic bridge was established due to the excellent conductively of ITO layer. If covering a sulfuric acid layer on the top surface of $WO_3$/ITO-glass, a protonic bridge was also established as shown in the schematic diagram of Fig. 1F. Based on this configuration, the migration of H atoms induced blue-color spread occurred in several seconds (Fig. 1C). This quick color change of $WO_3$ layer was also reflected by the detailed optical variation tests shown in Supplementary Fig. S4C. The observation was clearly demonstrated that the added electronic bridge could further accelerate the atomic migration if combined with the existed protonic bridge. The dynamic diffusion processes were visualized in the Supplementary Movie 1. From the different diffusional behaviors of H atom transport in $WO_3$ with and without cationic/electronic bridge (blue arrows in Supplementary Fig. S4), it was suggested that an intriguing mechanism existed behind this accelerated diffusion phenomenon.

## The polarized proton-electron synergistic diffusion

To understand the H atoms diffusion in $WO_3$ crystal film driven by the concentration gradient within the micro/nano scale size, we have fabricated an $H_xWO_3$-$WO_3$-$H_xWO_3$ hetero-junction on the deposited $WO_3$/sapphire film as shown in the optical image of Fig. 2A. The prepared $WO_3$ nanogap (as shown by the red circle) was about 5 μm, which was fabricated by the selected area hydrogenation based on the combination of UV lithography and metal-acid treated H doping process (see Methods). The topography image (Fig. 2B) mapped by scanning of Atom Force Microscopy (AFM) showed that the $WO_3$ nanogap had clear interfaces between the $H_xWO_3$ parts. The step analysis curve implied that the hydrogenation treatment would lead some corrosion of $WO_3$ layer, which made the $H_xWO_3$ session ~4.3 nm thinner than the intrinsic $WO_3$ area. The potential distribution of this $H_xWO_3$-$WO_3$-$H_xWO_3$ hetero-junction was also probed by Kelvin Probe Force Microscopy (KPFM) in Fig. 2C. It was observed that there existed clear potential difference between $H_xWO_3$ and $WO_3$ sessions and obviously the potential value of $WO_3$ was much lower. Since this potential value was closely associated with the work function of the tested materials, it was inferred that the $HxWO_3$ should have lower work function than that of $WO_3$ crystal.

This potential distribution was reasonable if considering the hydrogenation induced electron doping in $WO_3$ crystal, which would raise the Femi level ($E_F$) of $WO_3$. The theoretical calculation results (Fig. 2D and Supplementary Fig. S5) were also consistent with the above observations. In fact, from the projected densities of state (PDOS) of H-doped $WO_3$, it was clear that a transition from semi-conducting to metallic characteristic would be existed. This H doping induced continuous phase transitions were also verified by the resistance measurement in dynamic hydrogenation process via external voltage gating treatment, which showed that the $WO_3$ resistance decreased and transformed to metallic state gradually as the function of gating time (Supplementary Fig. S6).

In addition, due to the different work function at the $H_xWO_3$-$WO_3$ junction,

pronounced electron transfer from $H_xWO_3$ to $WO_3$ side would occurred. This electron transfer and the interfacial polarization were also confirmed by the first-principles calculations as shown in Fig. 2E. Though the electrons was easily transferred from $H_xWO_3$ to $WO_3$ side as shown in the scheme of Fig. 2F, the $H^+$ ions (protons) was quite difficult to move simultaneously even driven by the interfacial polarization, mainly due to the high diffusion barrier height in $WO_3$ lattice. Resultantly, the H atoms showed extremely low diffusion behavior in $WO_3$ crystal, as observed in Fig.1A or previous reports (*13-17*).

However, if adding a liquid acid solution layer covering the $H_xWO_3$-$WO_3$ junction, an effective $H^+$ bridge was quickly established via the solid-liquid interfaces. Accordingly, a polarized $H^+$-$e^-$ synergistic diffusion was formed for this system(*9*). It was suggested that at the first stage, the electrons transport from the $H_xWO_3$ side to pure $WO_3$ area under the driving force of work function difference (Fig. 2F). Then due to the charge redistribution, a localized interfacial polarization would drive the $H^+$ intercalating to the lower concentration area and extrapolating from the higher area via the $H^+$ bridge (acid solution), as shown in the scheme of Fig. 2G. Simultaneously, $H^+$ and $e^-$ recombination occurs in $WO_3$ part, resulting a quick in-plane hydrogen migration. Furthermore, if adding an electronic bridge (ITO layer was used) to improve the interfacial electron transfer process, the polarized $H^+$-$e^-$ synergistic diffusion would be further enhanced. It was pointed out that here the H diffusion in $WO_3$ films was obviously separated by protonic/electronic bridges at the solid-liquid interfaces, resulting an effective “job-share” diffusion. This proposed “job-share” diffusion greatly accelerated the H diffusion speed, which was quite consistent with the experimental observations for the quick color change in $WO_3$ layer within several minutes or even several seconds as shown in Fig.1A or 1B.

## Controllable ultrafast atomic diffusion in oxide

From the experimental observations, it was revealed that the “job-share” diffusion behavior greatly enhanced the H atoms diffusion in $WO_3$ layer due to the protonic/electronic bridges at the solid-liquid interfaces. In addition, the diffusion process should be closely associated with the conductivity of the protonic/electronic bridges respectively. Thus to gain more insight into the relationship between the diffusion rate and the protons/electrons conductivity, a detailed simulations by finite element analysis (FEA) was conducted to evaluate the diffusion behavior as the function of time and position (or the diffusion distance) in Fig. 3. According to the scheme of the $H_xWO_3$-$WO_3$ junction in Fig.3A, the H atoms diffusion was separated by the directional movements of protons and electrons via the polarized $H^+$ or $e^-$ bridges. This diffusion behavior could be described by the schematic circuit in Fig.3B, if considering the interfacial potential, the movement of electrons and protons through the $H^+$ or $e^-$ bridges based on *the Kirchhoff's lows*.

During the simulation, the hypothesis of linear relationship of potential *vs.* charge doping concentration was adopted (see Methods). To examine the effect of conductivity of e-bridge ($\sigma_e$) on the diffusion behavior during the "job-sharing" diffusion route, the conductivity of $H^+$ bridge was kept as a constant value. Then the simulated time-dependent distributions of the H atoms in $H_xWO_3$-$WO_3$ junction (Fig.3A) were plotted with the conductivity of $e^-$ bridge varied from $\sigma_e = 10^3$ to $10^5$ in Fig.3C~3E, respectively. The simulation results showed that H atoms would gradually diffuse from the $H_xWO_3$ side to undoped $WO_3$ side via the interface (position =0) and the H concentration showed a gradually decreased trend (from the position= -1 to 1). In addition, it was observed that the higher $e^-$ bridge conductivity would lead to the faster diffusion of H atoms. When the conductivity was increased to $\sigma_e = 10^5$, the H atoms in $H_xWO_3$-$WO_3$ junction could quickly go to a uniform distribution state as shown in Fig.3E, highlighting the important role of the conductivity of $e^-$ bridge. Fig.3F showed the H atom diffusion behavior as the function of diffusion time and the length (from position=0 to position=1) via the $H_xWO_3$-$WO_3$ junction if setting the H concentration to a constant value ($0.1C_{max}$). I was revealed that the diffusion time showed a quadratic-like relationship with the position when the conductivity of $e^-$ bridge was low (less than $10^3$). While when the conductivity was further increased to $\sigma_e = 10^4$ or $10^5$, the H atoms distribution could reach a balance state within much faster time. These simulations were quite consistent with the experimental observations in Fig.1C, which showed the quick color change in $WO_3$ layer on ITO substrate.

In fact the conductivity of $H^+$ bridge (or the $H^+$ concentration in sulfuric acid) also played important role in the "job-share" diffusion strategy. Since the color of $WO_3$ layer could be changed upon the H atom doping and the related transmission were directly lying on the H atom concentration, thus the visible transmission value of the $WO_3$ layer was able to be used as the indicator to evaluate the H diffusion behavior. In the experiment, we used sulfuric acid layer with different $H^+$ concentration (from $10^{-4}$ to $10^{-1}$ M) as the proton bridge and then examined the related transmission at different position as the function of diffusion time. Then the positional transmission evolution mapping of $H_xWO_3$-$WO_3$ junction was figured in Supplementary Fig. S7. If setting the transmission value of T=75%, the diffusion time as the function of diffusion distance (position from 0.0 to 0.4) was obtained in Fig. 4A .The experimental results revealed that the conductivity of proton bridge had the similar effect as the electron bridge during the H atoms diffusion in $WO_3$ layer. Based on these experimental results, the effective diffusion coefficients were obtained and revealed the positive relationship with proton concentration (Fig. 4B and Supplementary Table S3).

Furthermore, the ambipolar diffusion model(*30*) showed that $D_{eff} \propto \sigma_{H^+}\sigma_{e^-}/(\sigma_{H^+}+\sigma_{e^-})$, where $\sigma$ is the conductivity. Obviously, the $\sigma$ value would greatly increase due to the collective effects in protonic bridge(*16*). The additional high conductivity of e-bridge would further improve the effective diffusion coefficient if considering the "job-share" diffusion. Thus, accelerated diffusion process was simulated by the combination of proton and electron bridges. Five different $e^-$ conductivity values were considered in the diffusion coefficient calculation, which showed a positive relationship (Fig. 4C).

The effective diffusion coefficient $D_{eff}$ values for the current study and previous reports were listed in Table 1 for comparison. It showed that about four orders of magnitude of promotion for the $D_{eff}$ value could be achieved if applying the acid solution as protonic bridges. While if further combined the ITO as the e- bridge to form the “job-share” diffusion, another two orders of magnitude of promotion for the $D_{eff}$ value would be obtained as shown in Fig. 4D. This result showed that the $D_{eff}$ value could be controlled by adjusting the conductivities of the fabricated bridges for protons or electrons transport. More importantly, our current diffusion coefficient $D_{eff}$ values obtained through the “job-share” strategy were greatly increased up to ~ $10^{-1}$ $cm^2/s$, obviously exceed all of the previous reports completely, showing the overwhelming advantage.

## The universal of “job-share” diffusion behavior

According to the above ultrafast H atoms diffusion via the“job-share” mechanism, it was believed that this strategy was element independent and could be extended to other ions and materials(Fig. 5A). It was known that Li atoms intercalation and diffusion in electrode material were very important for the performance of Li-ion battery system. In the experiment, the accelerated diffusion of Li was also certified by the experimental results of $Li_xWO_3$-$WO_3$ junction, utilizing $LiClO_4$ dissolved in propylene carbonate (PC) as $Li^+$ bridges (Supplementary Fig. S8 and Supplementary movie 2).

Furthermore, it was also certified that this “job-share” diffusion strategy was suitable for another oxides. As a typical correlated oxide material, vanadium dioxide ($VO_2$) showed an insulator state at room temperature and had a typical metal-insulator phase transition property with the critical temperature of about 340K. While if doping some H atom into it, the $VO_2$ film would be stabilized in metallic state at room temperature. Accordingly, in the experiment, we just measured the film resistance distribution to monitor the H diffusion in $VO_2$ crystal lattice, which demonstrated that the H atoms diffusion in $VO_2$ crystal film was also greatly accelerated via the “job-share” diffusion route. From the surface resistance variation, it revealed that the quick hydrogenation induced metallization in the whole $VO_2$ film will be accomplished within several minutes (Supplementary Fig. S9).

For many practical device applications, an all-solid-structure was highly desirable to improve the mechanical stability and security. Thus, we proposed a structure designed for element fast storage and removal in bulk, as shown in Fig. 5B. In this structure, the catalyst, such as palladium (Pd), was utilized to split $H_2$ and inject H to the oxide. By this way, it was able to not only reduce the usage of expensive catalysts, but also make the structure more compact, comparing with classic device basing on films(*12*). To verify its advantages, prototype $WO_3$ film devices were fabricated in Fig.5C. Here the $WO_3$ films were deposited on conductive ITO substrates, some nano-size Pd islands were deposited and covered partial area to act as the catalysts. Typically of exposing this system in $H_2$ gas, the Pd-covered area would become blue color due to the H atoms doping into $WO_3$ layer. The uncovered area would still be transparent.

To fabricate an effective proton conductor, we just socked a solid polymer, polyacrylamide (PAAM) in sulfuric acid (1M) for 2 h, then a conductive proton conductor was obtained. Thus if we put this acid treated PAAM layer on $H_xWO_3$-$WO_3$ junction as a solid protonic bridge, the uncovered $WO_3$ area would transfer to blue color quickly and confirmed the important role of the solid protonic bridge to accelerate H diffusion (Supplementary movie 4). This solid protonic bridge would greatly benefit the practical device fabrication and integration. For comparison, the normal polymide layer also used to act as the protonic bridge. However, due to the poor conductivity, no H diffusion induced color change was observed for this system as shown in Fig.5C, further confirming the importance of the conductive solid protonic bridge.

## Conclusion and outlook

Achieving ultrafast atom diffusion in functional solids had been a challenge for many years. In the current study, we achieved an accelerated diffusion of doping atoms in oxide films via a synergistic job-share strategy, which separated the electronic and cationic transport pathways by configuring the artificial electrons/cations bridges.

Based on this polarized cation-electron synergistic diffusion route, an ultra-fast diffusion of H atom in $WO_3$ was directly visualized with the $D_{eff}$ value up to ~ $10^{-1}$ $cm^2/s$, which was about six orders of magnitude higher than that in the traditional $WO_3$ bulk and overwhelming the previous reports. This cation-electron synergistic diffusion was general and universal, which would be applicable for other doping elements and oxide systems. It was believed that the ultrafast atom diffusion would potentially stimulate various frontier research areas, such as ultra-fast charge batteries, high efficiency catalyst, neural network or other emerging ionic devices.

**Acknowledgements**

This work was partially supported by the National Natural Science Foundation of China (12074356, 52130601), the Fundamental Research Funds for the Central Universities, the Open Research Fund of State Key Laboratory of Pulsed Power Laser Technology and the Youth Innovation Promotion Association CAS. This work was partially carried out at the USTC Center for Micro and Nanoscale Research and Fabrication. The authors also acknowledged the supports from the Anhui Laboratory of Advanced Photon Science and Technology. L. L., J.J. and C.Z. conceived the study. L. L. and C.Z. designed the experiment. M.H. and J.J. conducted the theoretical calculations. L. L., C.H., B.L., S.Z., G.Z. and L.B. L. conducted the initial tests. L. L., J.J., C.Z. and L.B. L. wrote the manuscript. All authors discussed the results and commented on the manuscript.

**Supplementary Materials**

Materials and Methods

Figure. S1 to S10

Table. S1 to S4

References

# Figures

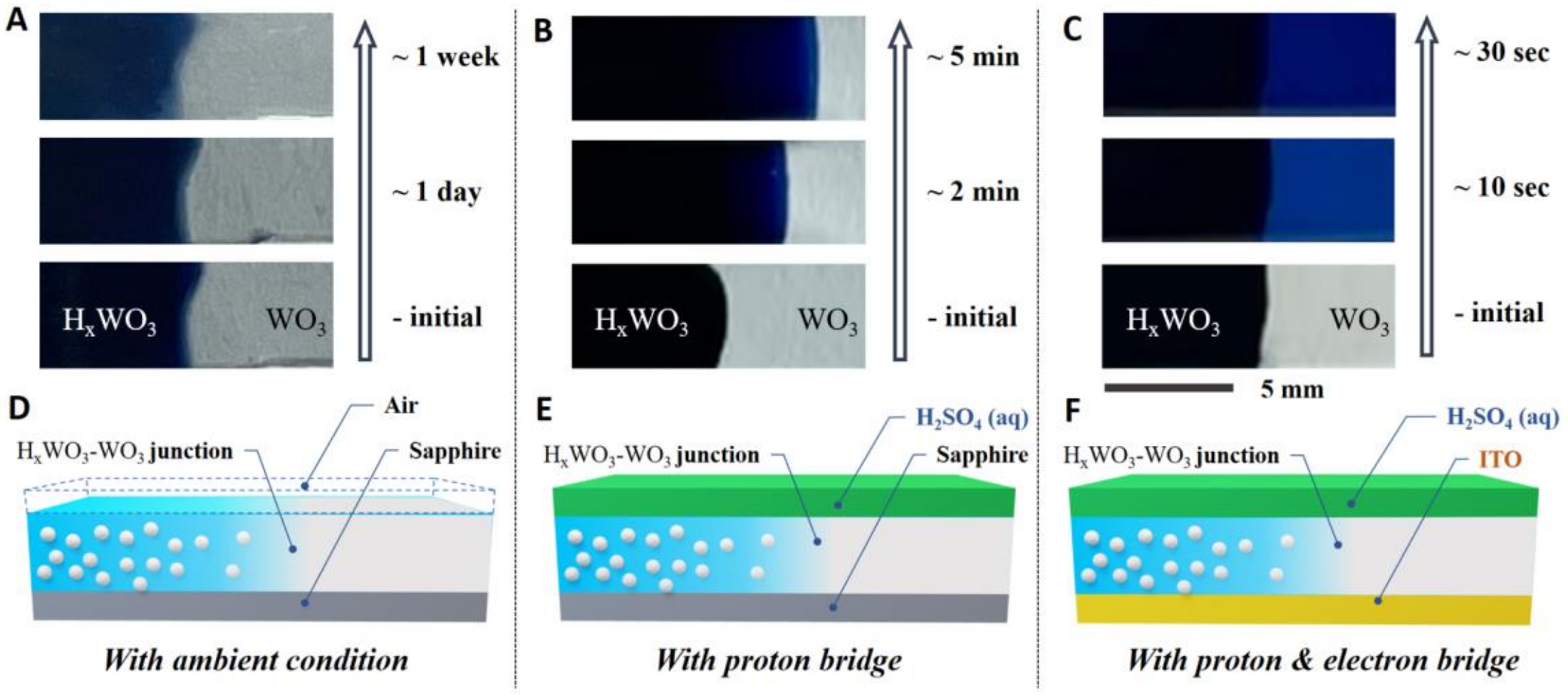

**Fig. 1. Accelerated diffusion of hydrogenation atoms in $WO_3$ films.** (**A** to **C**) Optical images for the H diffusion in $WO_3$ films under different conditions, which could be directly visualized due to the electrochromic feature of H-doped $WO_3$. Three different sandwich structures were conducted for the experiments, showing different diffusion time. (**D** to **F**) The schemes for the diffusion behaviors of hydrogenation atoms in $WO_3$ films under different conditions.

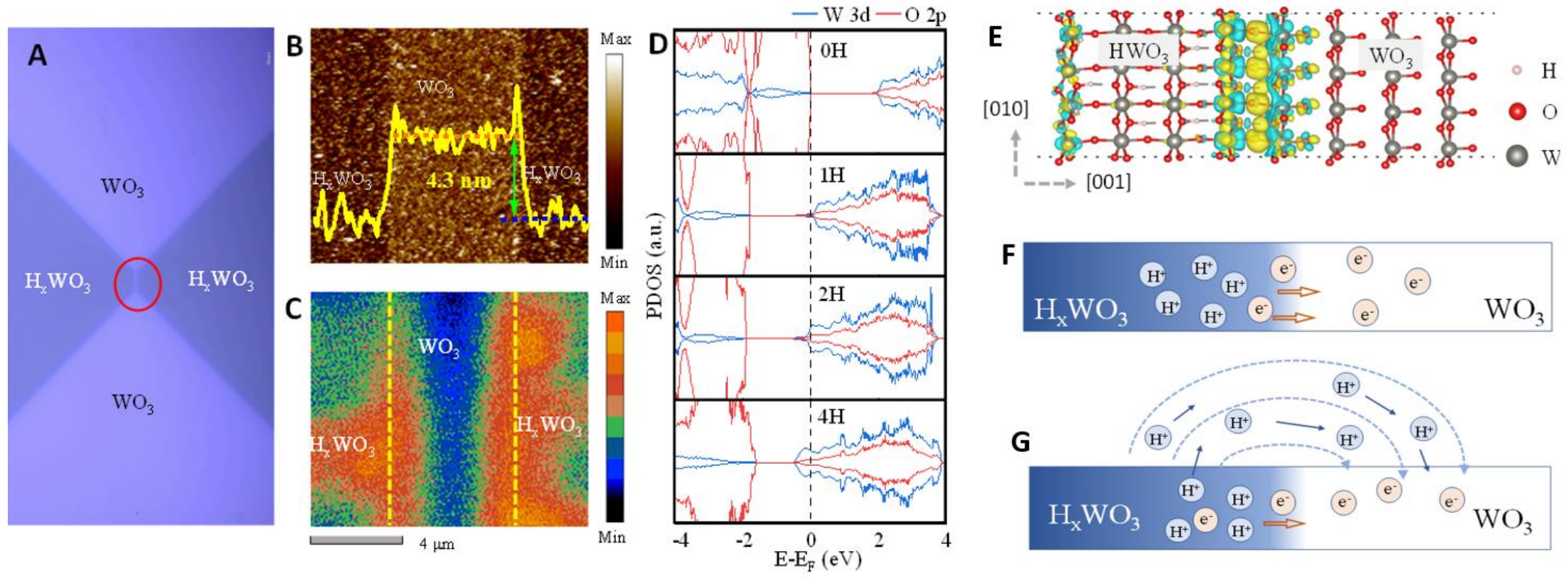

**Fig. 2. The mechanism of proton-electron synergistic diffusion**. (**A**) The optical micro-image for the planar structure of hydrogenated-intrinsic-hydrogenated $WO_3$. (**B**) The atomic force microscope (AFM) image for the selected nano-gap area marked by red circle in (A). The yellow solid line showed the line-scan for the gap, which shows that the intrinsic area is~4.3 nm higher than the hydrogenated area. (**C**) The Kelvin probe force microscope (KPFM) measurement for the same nano-gap in (B), showing the boundaries of the contact potential. (**D**) The calculated projected density of states (PDOS) of $WO_3$ with different H-concentration. (**E**) Charge difference at the interface between $HxWO_3(001)$ and $WO_3(001)$. The value of the iso-surface was 0.0026 e/Å$^3$. Yellow and blue iso-surfaces indicated the accumulation and depletion of charge density. (**F)** The scheme for the interfacial electron transfer from $HxWO_3$ to $WO_3$ side due to the different work function. (**G**) The schematic process for $H^+$ directional movement through the proton bridge driven by the solid-liquid interfacial potential.

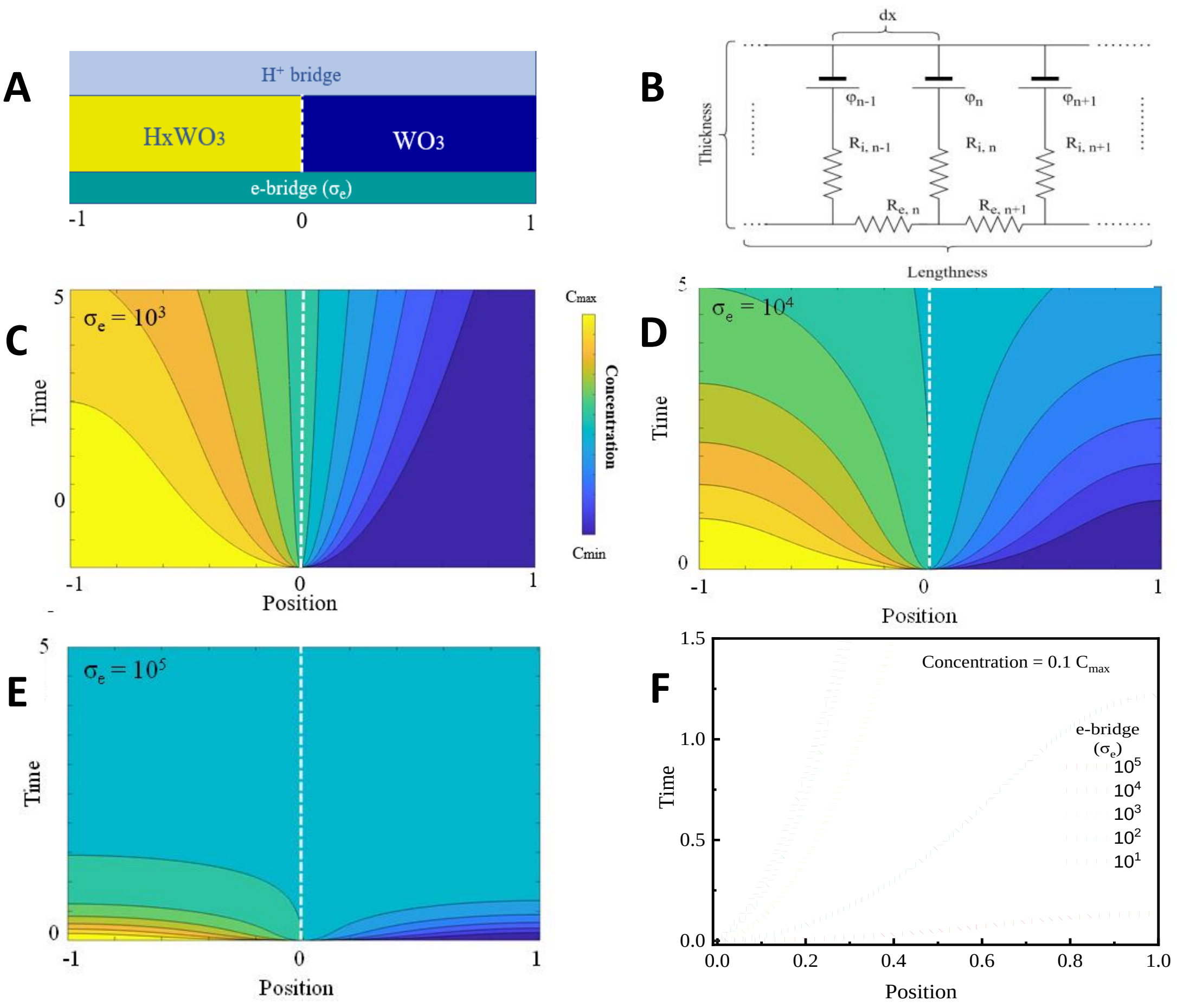

**Fig. 3. The simulations of accelerated diffusion with the method of finite element analysis (FEA).** (**A**) The corresponding model and initial conditions of this simulation. (**B)** Schematic of one-dimension resistor network to simulate the diffusion process. (**C~E**) the results of positional concentration evolution, with different conductivities ($\sigma_e$) of the substrate, from $10^3$ to $10^5$ a.u.. (**F**) The relationship between the diffusion time and diffusion position (from 0 to 1) under different conductivity of electron bridge if setting the H concentration to be a constant value (0.1 $C_{max}$).

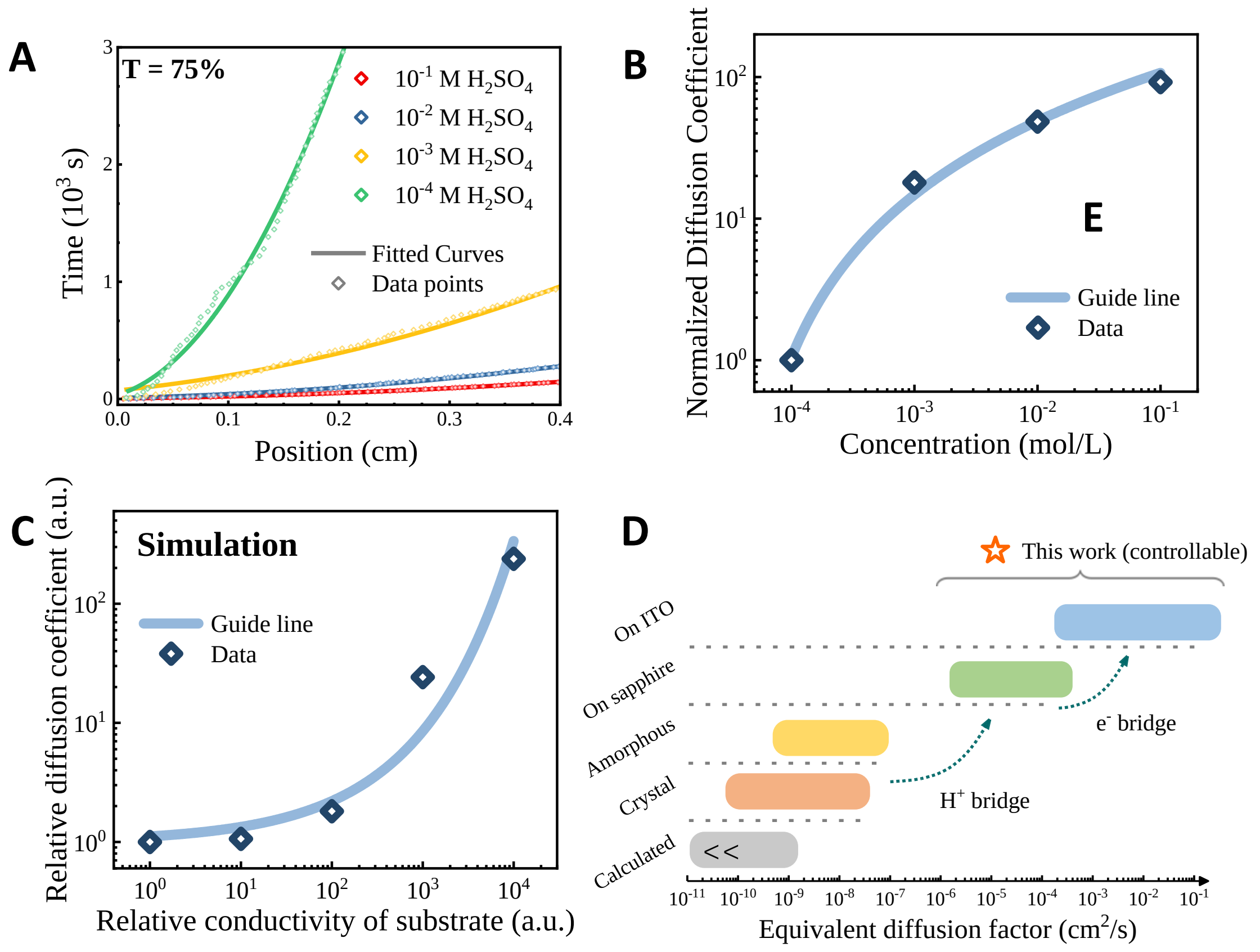

**Fig. 4. The dependence of protonic and electronic bridges in accelerated diffusion.** (**A**) For the $H_xWO_3$-$WO_3$ junction with the cover of sulfuric acid as the protonic bridge, the relationship between the diffusion length (borderline position) and the time under different concentration of sulfuric acid, from $10^{-4}$ M to $10^{-1}$ M. The borderline position was confirmed according to the visible transmittance of T~75 %, which was extracted from the transmission-position curves as shown in Supplementary Fig. S4B. (**B**)The increasing trend of normalized diffusion coefficient, $D_n = D_{eff}/D_{Min}$ as the function of acid concentration (supplementary Table S3). (**C**)The normalized diffusion coefficient $D_n$ as the function of electron conductivity of substrates, which were obtained from the simulation results in Supplementary Fig. S7. (**D**) The comparison of the effective diffusion coefficient $D_{eff}$ values shown in Table 1. It showed that ~$10^4$ times of promotion for the $D_{eff}$ value with the $H^+$ bridges. If further combined the $e^-$ bridges, another ~$10^2$ times of promotion would be obtained with the $D_{eff}$ value of up to ~ $10^{-1}$ cm$^2$/s, overwhelming all of the previous reports.

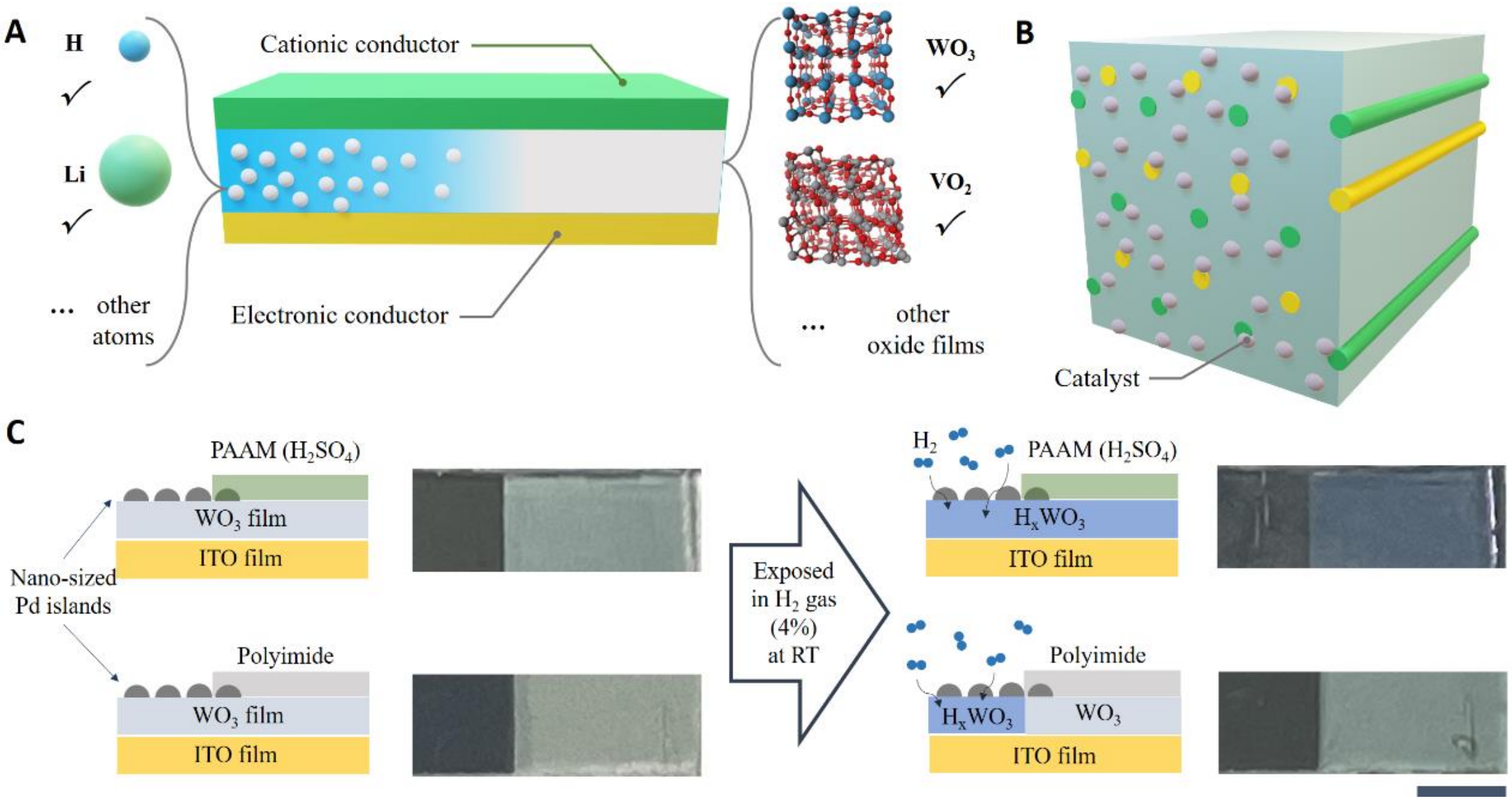

**Fig. 5. The feasibility in other system with ultrafast hydrogen storage/removal.** (**A**) The schematic diagram for this cation-electron synergistic accelerated diffusion, which was also suitable for other elements (such as Li) or other oxides (such as $VO_2$). (**B**)The proposed bulk structure designed for ultra-fast hydrogen (or other eligible elements) storage and removal system. (**C**)Top row, the prototype $WO_3$ films device for hydrogen storage: Nano-sized palladium (Pd) was deposited on a selected area and the $H_2SO_4$ (0.1 M) treated polyacrylamide (PAAM) film were selected as the protonic bridges. Then, the whole $WO_3$ film was converted to blue color, which could be visualized in Movie 4. Bottom row, the comparative experiment without a protonic bridge: only the exposed $WO_3$ area became blue color. The scale bar for the optical images was 5 mm.

**Table 1:** The effective diffusion coefficient of H atom in $WO_3$.

| Samples | $D_{eff}$ $cm^2s^{-1}$ | Ref |
|---|---|---|
| $H_2SO_4$(1E-1 M)/$WO_3$/ITO | $\sim 1.63\times 10^{-1}$ | This Work |
| $H_2SO_4$(1E-1 M)/$WO_3$/$Al_2O_3$ | $\sim 1.01\times 10^{-4}$ | This Work |
| a-$WO_3$ | $2.90\times 10^{-10}\sim 1.33\times 10^{-8}$ | Ref (*36*) |
| a-$WO_3$ | $5.10\times 10^{-10}\sim 3.45\times 10^{-9}$ | Ref (*34, 37*) |
| c-$WO_3$ | $2.70\times 10^{-9}\sim 3.28\times 10^{-8}$ | Ref (*36*) |
| $WO_3$ | $4\times 10^{-10}$ | Ref (*38*) |
| c-$WO_3$ | $2.9\times 10^{-11}\sim 7.1\times 10^{-11}$ | Ref (*39*) |
| $WO_3$(calculated) | $<10^{-9}$ | Ref (*2, 40, 41*) |